\documentstyle[epsfig]{elsart} 
\begin{document} 
\topmargin=-1.0cm  
\oddsidemargin=1.0cm                                            
\begin{frontmatter} 
 
\title{Sliding Objects with Random Friction} 
\author{Itzhak Webman, Jos\'e Luis Gruver, and Shlomo  
Havlin} 
\address{Department of Physics and Jack and Pearl  
Resnick Institute of  
Advanced Technology, Bar-Ilan University, Ramat-Gan  
52900, Israel} 
\maketitle

\begin{abstract} 
 
We study the motion of elastic networks 
driven over a random substrate. Our model which  
includes local friction forces  leads to complex  
dynamical behavior. We find a transition to a sliding  
state which belongs to a new universality class. The  
phase diagram comprises of a pinned state,  
a stick-slip motion phase, and a free motion phase. 
\end{abstract} 
%\pacs{PACS numbers: 46.10.+z 64.75.+g 81.35.+k} 
%\linespread{1.6} 
%section{Introduction} 
%\linespread{1.6} 
 
\end{frontmatter} 
 
\section{Introduction} 

The dynamics of elastic objects  
driven  through an environment with random pinning  
interactions is complex and its present  understanding    
still incomplete. Processes of this type occur in   
microscopic systems such as charge density waves  
driven by an  electric field \cite{CDW,FLR}, 
superconductors  subject to external magnetic  fields  
\cite{LRK}, and polymers moving through a  
inhomogeneous environment. Related  processes  on a  
macroscopic scale  occur in the motion of geological  
faults \cite{BK,CL,FDR}, or  in sliding friction between 
an elastic object and a rigid one \cite{VG}.  Theoretical  
studies of these phenomena are usually modeled  
by  an elastic network moving  in a rigid environment 
or over a rigid substrate with random interactions  
between them. In the simplified case where  one of 
the two objects is homogeneous, one  may distinguish 
between  $\it random\;network$ models, where the  
random interactions are assigned to sites on the elastic  
network while the medium or substrate are  
homogeneous, and the $\it random \;  
substrate $, models where the random interactions are  
assigned to the sites of the substrate or the embedding  
medium while the elastic network is homogeneous.  
These  two type of models  are illustrated in Fig. 1  
 
Random network models were  invoked to  describe  
randomly pinned charge density waves  (CDW)  in  
solids \cite{DF}-\cite{P} and were widely studied.    
Most recent work  on random substrate models  
concentrated on the high velocity limit, where the  
distortions of the network induced by the random  
interactions are mild \cite{GLD}-\cite{BAL}, while the  
dynamical properties near the critical external field for  
depinning received less attention.   
We focus on the simplest  $\it discrete$  realization of  
the random substrate model containing no ad-hoc  
features such as velocity dependent forces \cite{CL}, and  
compare it to the analogous random network model.  The  
study shows  that the dynamics of the random substrate  
model above depinning threshold is interesting and  
 complex,  featuring several distinct regimes.  
In contrast, in  the random network case  the  elastic  
object  follows a simple steady state motion. 
The  pinned state of the random substrate model below  
threshold is characterized by a wide distribution of $\it  
strain\; avalanches$.  Similar avalanches occur in the  
random network model, however the critical exponents  
describing the distributions of avalanche sizes are  
definitely different.  
 
The properties of the discrete  random substrate model  
depend on three parameters: The external field, the ratio  
of pinning strength to the elastic stiffness,  and the  
characteristic size of  the local pinning regions.  We find  
that whereas  in some parts  of the relevant phase  
diagram the two models are approximately equivalent,   
through most of it there are pronounced differences  
between them, and the critical behavior belongs to  
distinct universality classes. 

\section{The Model} 

Sliding elastic objects are represented here  by a discrete  
chain of particles connected by springs. (see Fig. 1)  
In the random network limit each particle is attributed a   
random  pinning force, while the substrate is  
homogeneous.  
For an external force  $E$ switched on at $t=0$ acting  
uniformly on all particles, the dynamics of the   
displacements  of the positions of the particles from their  
initial equilibrium  positions: $\xi_i = x_i(t)-x_i(0)$ is  
described by the following overdamped equations of  
motion:   
 
$\it random\;network$ 
\begin{equation} 
{\partial{\xi_i}}/{\partial{t}}=  
\max(0,\kappa({\xi}_{i+1}+\xi_{i-1}-2\xi_i)+E- 
V(i)) 
\end{equation}  
This realization of the random network model, also  
known as the $\it ratchet\;model$  
\cite{W87} successfully describes the dynamical  phase   
transition of charge density waves  from a pinned state to  
a DC current  conducting state  at a critical external field   
$E_c$ \cite{W87,puri,MF,MS,FLRM}.  
 
In the random substrate model the particles interact with  
the substrate via local pinning forces $ V(x_i)$ ,  
whereby the i'th  particle at position  $x_i $ (in the  
substrate frame of reference) moves if the total force  
acting on it is greater than $V(x_i)$. As before, the  
dynamics for the  displacements: $\xi_i = 
x_i-{x_i}(0) $, is described by the following equations: 
 
$\it random\;substrate$ 
\begin{equation} 
{\partial{\xi_i}}/{\partial{t}}=  
\max(0,\kappa({\xi}_{i+1}+\xi_{i-1}-2\xi_i)+E- 
V({\xi_i}+i)) 
\end{equation} 
Initially, the particles  positions form a  1D lattice with  
unit spacing:  
${x_i}(0)=i$.  
Pinning forces are local, with a characteristic range  
$\Delta $.   In our  study the  pinning force is constant  
inside lattice unit cells, ($\Delta=1 $), and  
varies randomly from one cell to the next. 
We chose a binary  distribution for $V$ of the type:  
$\rho(V) = p\delta(V-V_1)+(1-p) 
\delta(V-V_0)$, usually with p=0.5.  
We adopt periodic  boundary conditions, so  that the  
argument of $V$ in equation (1) is  $  
{\xi_i\;mod}(L)+i $.

Both models  also  depict a discrete one  dimensional  
elastic interface moving  in the $\xi$ direction in a $ \it  
two\;dimensional$ $(i,\xi)$  plane \cite{puri}, where 
the interface is initially $\it flat$.   
In the random substrate  model a  pinning  $\it site$ on   
the $i$ axis,  
transforms into lines of pinning sites of slope unity in the  
$(i,\xi)$  plane, as described in Fig. 2.  
 In the random network case, pinning sites form  
$\it columns$  of fixed  pinning strength $V_i$ in the  
same plane.   
 
$\it Approximate\;scaling\;properties$ 
  
The  dependence of the general behavior of the random  
substrate model on the elastic  
stiffness $\kappa$, and  on the  size $\Delta$ of regions  
where pinning forces do not vary, can be simplified by  
the following scaling relation \cite{valid} (provided  
$\max(\kappa,\kappa\Delta)<1$):  
\begin{equation} 
\xi(\kappa,\Delta, t)= {\kappa^{-1}}\xi(1,\kappa \Delta,  
t/\kappa ) 
\end{equation} 
 
Extending this scaling relation to  $\Delta >1$  results in  
a hybrid random network-random substrate model, so  
that  the pure random substrate model is not preserved.  
However, in  the limit  $\Delta>>1$  the pure random  
network model is approached, so that the random  
network model and the random substrate model approach  
each other in the very high stiffness limit, as $\kappa  
\rightarrow\infty$.

\section{Dynamical properties}   
 
In the random network model above the   
threshold  $E_c$ all  particles eventually move with the  
same  finite velocity as  a $ \it  rigid$ distorted  object.  
Adding  the $N$ equations of motion Eq. (1)  the elastic  
forces sum-up to zero, so that the velocity of  
the center of mass is:  
\cite{W87} 
 
\begin{equation} 
{U_{RN}}=E-E_c 
\end{equation} 
where for a chain of $N$ particles $E_c={1\over  
N}{\sum V_i}$

The dynamical behavior  of the {\it random substrate}   
model  is more complex,  and   two  distinct regimes  
appear as the driving force is varied above the depinning  
threshold $E_1$:  
 
$\it  stick-slip\; regime$ : $E_1< E < E_2$ ;  
 
The center of mass of the  system moves with a roughly  
constant velocity,  modulated by fast  fluctuations,  
although at any time $ \it a\; finite\; fraction$ 
of the particles are not in motion. 
 
The center of mass  velocity $U$ obtained numerically 
for several values of stiffness $\kappa$ is shown in Fig. 3. 
Each curve corresponds to a single representative  
random configuration for chains N=100-1000.   
At the threshold there is a small discontinuity, followed  
by a linear $E$ dependence.  
The fractions  of particles that are in motion  
at a given instant  corresponding to the velocity curves in  
Fig 3. appear in the inset.   
 
Deriving the CM velocity by summing the individual  
equations Eq. (2) is more subtle  in the random substrate  
case, since the equations  for particles with  zero velocity  
are actually $\it  inequalities$.    Time averages  over  
sufficiently long  periods are equivalent to those obtained  
from a linearized  version of  Eq. (1,2) where the  ratchet  
condition is omittted, leading to: 
\begin{equation} 
{U_{RS}}= E -\sum n(V_i,E) V_i 
\end{equation} 
where $n(V_i,E)$ is the time averaged  fraction of  
particles occupying the i'th unit cell.  Since the residence  
time of particles over a strong pinning  area is longer  
than over a  weak pinning  one, the mean  occupation  
fractions $n(V_i,E)$ are monotonously increasing  
functions on $ V_i$ , so that  for  two  systems sharing  
the same distribution of random pinning strengths,  the   
mean velocity of the random network is always greater  
than that of the random substrate system,  asymptotically  
approaching  it  from below in the limit of high velocity.  
The threshold $E_1$ is the solution of the equation:  
\begin{equation} 
E_1 -\sum n(V_i,E_1) V_i 
\end{equation} 
so that the inequality $E_1>E_c$ is always valid.  
 
The center of mass velocity of the random substrate  
model  can be expressed as a driving force minus a  
velocity dependent drag  force $F(U)$: 
\begin{equation} 
U_{RS}=E-F(U_{RS}) 
\end{equation} 
Analysis of the velocity data shows that $F(U)$ is equal to the  
static friction force $E_1$ in the limit of zero velocity,  
{\it increases}  with increasing velocity above  $E_1$  reaching some  
plateau, eventually tending to $E_c$ for high velocities.
This  weakening of the kinetic friction force with increasing 
velocity  is  generated  by  a detachment instability where the 
fraction of immobile particles abruptly  vanishes (Fig. 3 inset).
Qualitatively similar features were observed in experiments 
where an elastic membrane is dragged over a 
rigid substrate, performing stick-slip motion. \cite{VG}
 
The center of mass velocity  for $E$ just above the  
depinning transition scales as  $(E-E_1)^\beta $.    For  
the random network model, trivially, $\beta=1$.  Our  
results for the random substrate model also follow a  
linear $E-E_1$ dependence, i.e. $\beta=1$. This is in  
contrast with Ref. \cite{CH}, where  $\beta =0.47$ is  
reported. The differences in exponents may be due to   
differences between our model and the specific   
realization  of the random substrate model in  
Ref.\cite{CH}.  This is supported by the fact that   
in nonlinear dynamical systems with quenched    
randomness, universality is often weaker than in   
equilibrium critical phenomena, so that some critical  
exponents  may depend on details on short length scales.     
 A striking demonstration of this for a directed  
polymer in a random medium, a problem closely related  
to the one considered here, was published recently  
\cite{LZ}.  
  
$\it  free\; motion\; regime$: $E> E_2$ ;  
 
In this regime  all particles possess a finite velocity at all  
times (See Fig. 3, inset). The instantaneous velocity of  
the center of mass is made up of a constant part plus  a  
random $\it washboard $ like  modulation induced by  
motion over a fluctuating pinning landscape. In the limit:  
$E>>E_2$, the velocity tends to that of the analogous random  
network Eq. (3),  
and the relative fluctuating component of the velocity  
diminishes (Fig. 3). The drag force $F(U)$ approaches  
$E_c$. 
  
Our model assumes the  presence of  local  friction  
forces randomly distributed along the interface, while  
their microscopic origin are outside its scope.   
The  stick-slip dynamics in the random substrate model   
is a consequence of the random arrangement of simple  
local interactions together with the cooperative effect of  
the elastic forces.   While the  decrease of macroscopic  
kinetic friction with velocity in real materials may  have  
various causes, comparison with our results suggests   
surface inhomogeneity or randomness may play a  
significant role.  
 
\section{Scaling Properties of the Pinned Sate} 
 
For both models, interfaces evolving by the dynamics  
described in Eq (1) or Eq (2), starting from a flat initial  
configuration,  ($\xi_i=0$ at t=0) and subject to a  
constant drive $ E $  below the pinning threshold,  
eventually reach a static state of strongly strained  
domains, or $\it  
strain\;avalanches$ 
\cite{PIN}, seperated by  $ \it virgin $ particles which  
never moved. For $E$ approaching $E_c$ from below,  
the random network model undergoes  a  second order  
dynamical phase transition \cite{W87,puri}.  
The following numerical results show that the random  
substrate model follows a qualitatively similar critical  
behavior, but of a different universality class.  
  
Fig. 2 shows a typical pinned 
interface for  
the random substrate model. 
Here interfaces are hindered  by  
the tilted  pinning lines, leading to characteristic  
triangular structures with a roughness exponent  
$\zeta=1$. In contrast, in the 
random network  model, distortion of domains is  
much more pronounced and the roughness exponent is  
$\zeta=3/2$  
\cite{W87,puri}. 
Similar to the random network model and to many other 
non-equilibrium systems (e.g. the scaling of avalanche  
sizes in self-organized criticality: \cite{BTW}),  
the  $\it number $ of domains of size $\ell$  per  
unit length in the random substrate model has the  
generic scaling form:  
 
\begin{equation} 
n(\ell)\sim {\lambda^{\sigma}}{{x}^{-\tau}}  
\Phi(x)     
\end{equation} 
where  $x=\ell/{\lambda(\epsilon)}$, and $\Phi(x)$ is a  
slowly varying function for $x<1$ with  
a steep  cutoff at $ x>1$. The typical size of the  
largest domains, $\lambda$,  diverges as  
$\epsilon^{-\nu}$   where $ \epsilon=E-E_1 $.  The  
number of domains is not fixed.  Close to the depinning  
threshold the number of virgin particles  vanishes, and  
the whole system is tiled by strained domains.  This  
global condition  implies that the first  moment of $n(l)$  
is a  slowly varying function of $\epsilon$  that does not  
become singular at criticality, leading to $ {\sigma}=2 $ 
for  $\tau>1$, while for  $\tau<1$, $\sigma=1-\tau $. 
The exponents of the random network model  
are different: $ \tau=3/2$, $\sigma=2$,  $\nu=2$  
\cite{W87}.  
   
We obtained $n(\ell)$  for the random substrate model by   
solving Eq (1)   numerically for   an ensemble of  40  
random realizations of chains of $10^4$  particles.  Fig.  
4  shows the rescaled   histograms. The collapse of the   
data for  different values of $\epsilon$ was achieved  by  
rescaling  using $\nu = 2 $ and  $\tau=0.9 \pm 0.1$.   
 
The analysis of higher moments of  $ n(\ell) $ does not  
show significant deviations from single parameter  
scaling.  The  values of critical exponents are  
siginificantly different than those of the random network  
model, confirming  that the two  models belong to  
distinct universality classes.   
\section{The Phase Diagram} 
 
The properties of the  random substrate model depend on  
the  network's elastic stiffness. Intuition suggests that  
when the network becomes very rigid,  the differences  
between the two systems should vanish.   
Fig. 5 represents a ($\kappa\;  E$) phase diagram,  
obtained  from numerical studies of chains of N=100.  
As the stiffness increases, the random  
substrate threshold $E_1$ approaches $E_c$.  The  
second threshold  $E_2$ is of the order of $V_{max}$, the  
upper bound of the pinning strength distribution, and depends only  
weakly on $\kappa$.  
   
Within the pinned regimes of both models, in region I of  
the phase diagram, configurations of strained domains  of  
both models are roughly the same.  The dark circles  
denotes values of $\kappa, E$ where the relative  
Hamming distance $D$ \cite{HAM} between locations   
of strained domains for pairs of  systems from each   
model with the same set of pinning  strengths $\{V_{i}\}$,  
is smaller than 0.06.       
 
For the random network model the largest displacement 
${\xi_{max}}(\epsilon)$ is bounded by  
$\lambda(\epsilon)^{3/2}\sim \epsilon^{-3}$. 
The scaling  of the random substrate model with stiffness  
given by Eq (3), implies that for large  
$\kappa > \xi_{max} $ the two models track.    
This is a sufficient condition , so it yields  a lower  bound 
for the boundary of  region I  where the  
two  models coincide, given by values of $\kappa,E$:  
\begin{equation} 
E_c-E \sim {1\over \kappa^{1/3}}  
\end{equation} 
An  upper bound for  $E(\kappa)$ is given by the relation:
 $E_c-E \sim 1/\kappa$ \cite{valid}.
The corresponding boundary is also shown in Fig. 5. 
 The fact that for stiff networks  the two models are roughly 
equivalent, means that  aspects of friction between very stiff 
solid bodies 
which are  related to inhomogeneities on the interfaces can be  
described by the random network model,  or by models  
based on it, elaborated to better conform to real solids.

\vfill\break  
%{\bf Figure Captions} 

\begin{figure} 
\epsfig{file=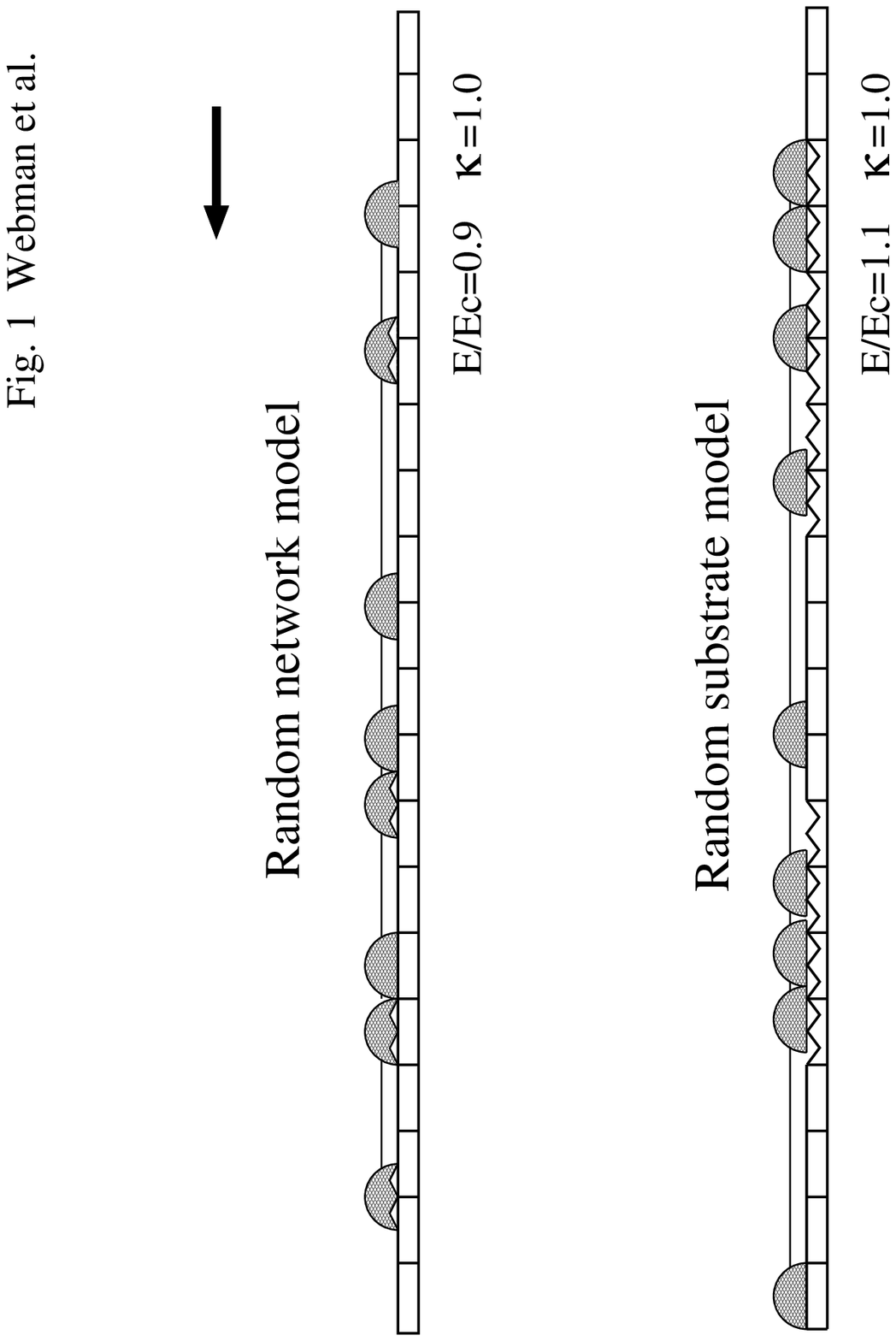,width=10cm,clip=,bbllx=200,bblly=140,bburx=550,bbury=650,angle=-90}
\caption{%Figure 1:
Illustration of the random network and the  
random substrate models.}
\end{figure}
 
\begin{figure} 
\epsfig{file=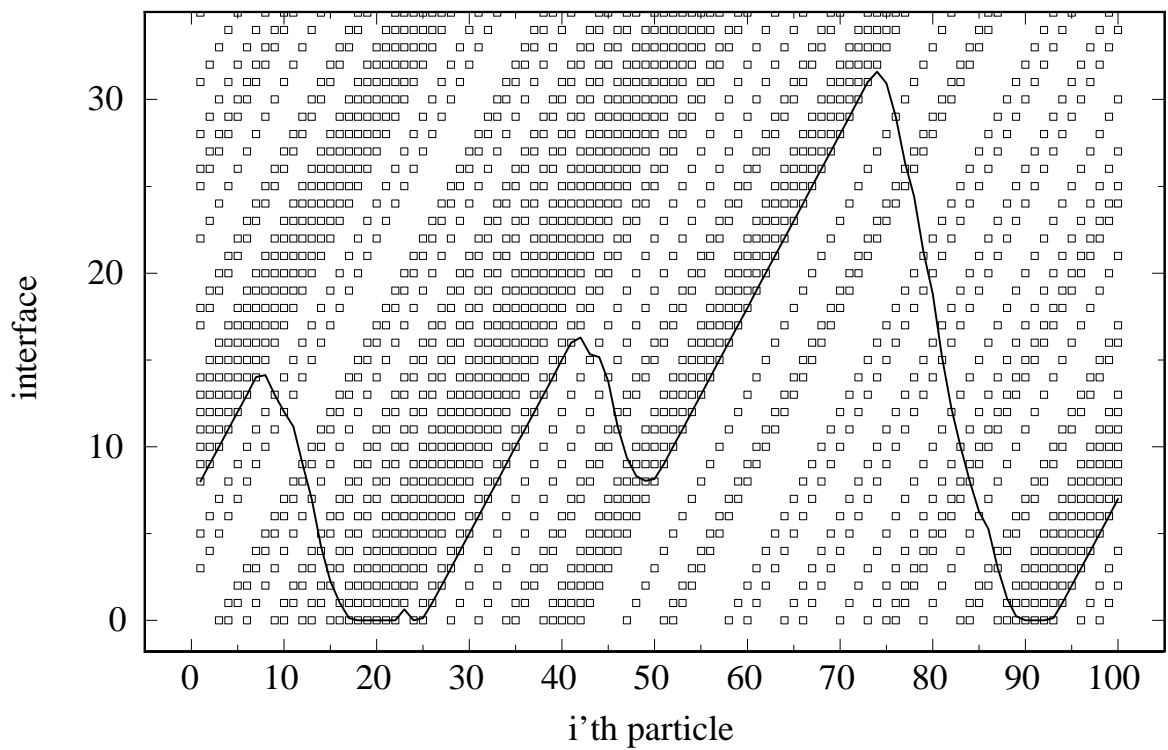,width=10cm,clip=,bbllx=165,bblly=105,bburx=535,bbury=680,angle=-90}
\caption{%Figure 2:
A charactersitic pinned  configuration 
in the interface representation of the random substrate  
model.}
\end{figure}
  
\begin{figure} 
\epsfig{file=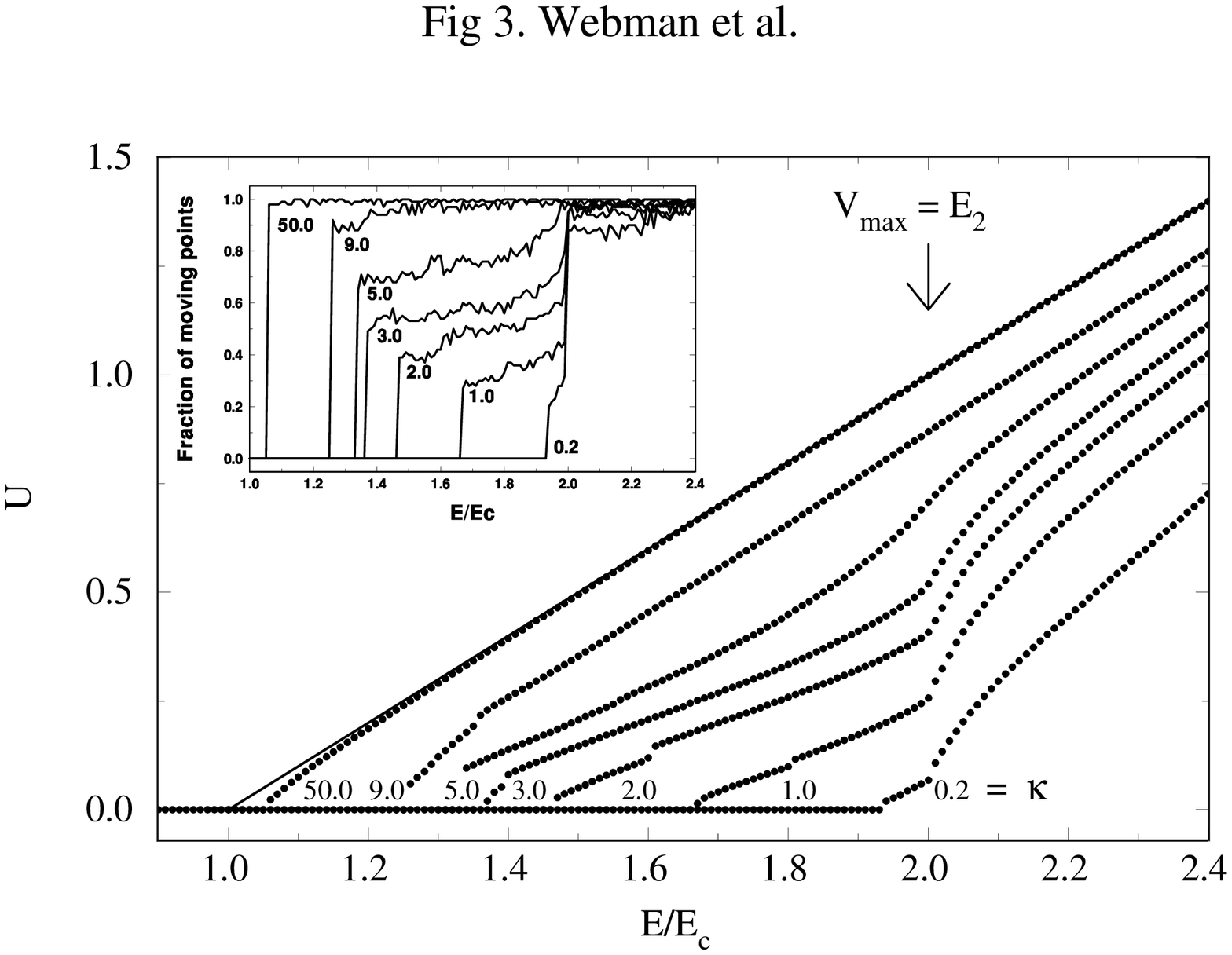,width=10cm,clip=,bbllx=160,bblly=110,bburx=540,bbury=670,angle=-90}
\caption{%Figure 3:
Time averaged center of mass velocity vs. 
driving force for networks of various stiffness. 
Inset: Mean fraction of particles in motion.}
\end{figure}
 
\begin{figure} 
\epsfig{file=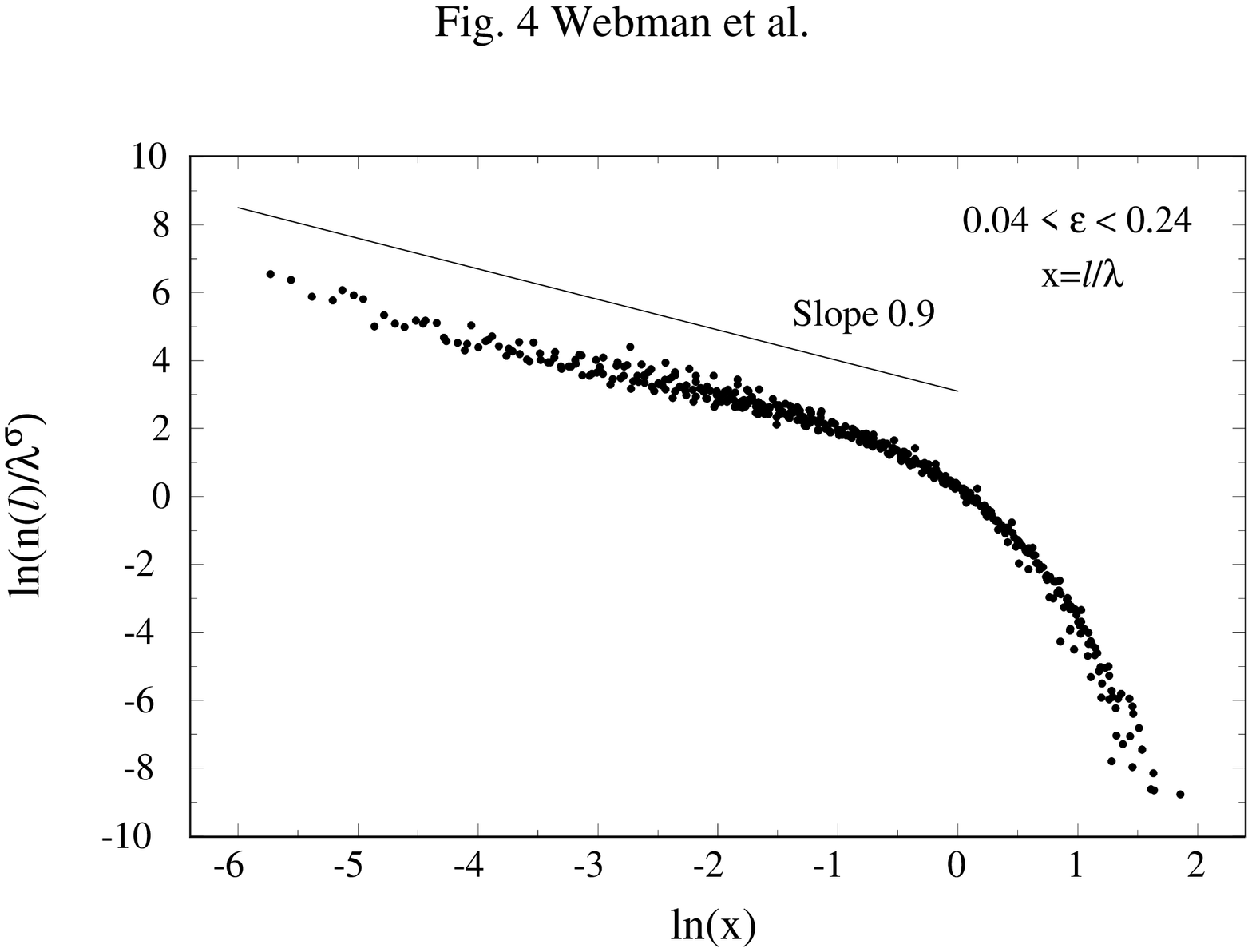,width=10cm,clip=,bbllx=160,bblly=100,bburx=540,bbury=680,angle=-90}
\caption{%Figure 4:
Log-log plot of scaled distribution of sizes of  
strain avalanches in the pinned regime.}
\end{figure}
 
\begin{figure} 
\epsfig{file=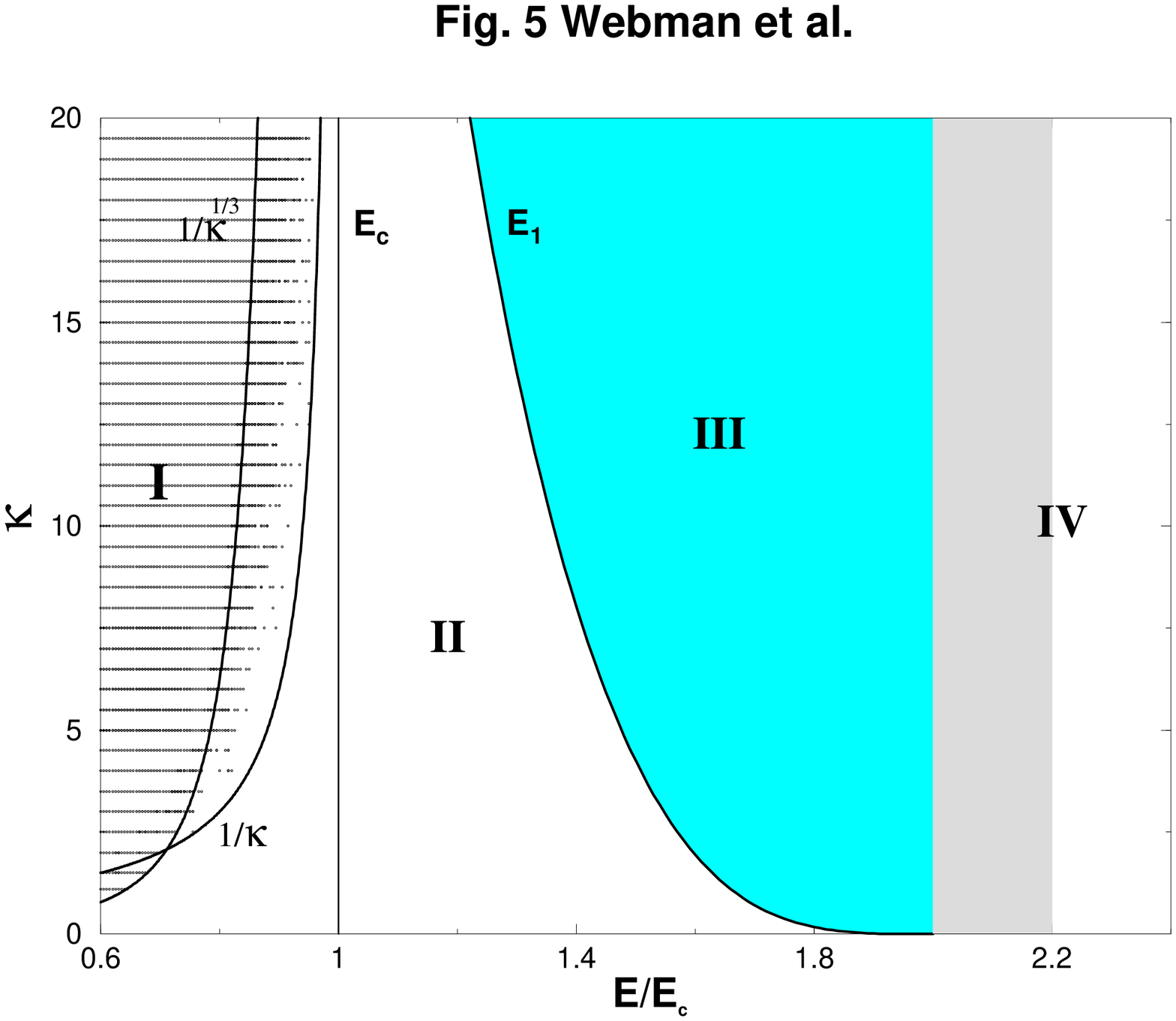,width=10cm,clip=,bbllx=95,bblly=80,bburx=550,bbury=660,angle=-90}
\caption{%Figure 5:
Phase diagram in the  $(\kappa, E)$.
The network configurations are pinned in phase I and phase II.
Phase III corresponds to stick-slip dynamics, and phase IV 
to free motion. In  phase I the random network and the random 
substrate  models  yield  similar pinned configurations. 
Theoretical lower and upper bounds of the boundary of phase I 
are  shown.}
\end{figure}


\begin{thebibliography}{99}
\bibitem{FLRM} In the  FLR-Fisher equation is  
similar to Eq(2), except that  there $V(i)$ is multiplied by a   
cosine function of $\xi $.  Using the ratchet condition   
is similar to  replacing the cosine force by a derivative of a 
sawtooth potential,  except that local backword motions down 
the slope of  the sawtooth are excluded. These motions  do not 
affect the large  scale dynamics  since the model has no inertia. 
The  critical behavior of the random network model below 
threshold is in the  same universality class as  the  CDW  
models   of Fukuyama Lee and Rice and Fisher \cite{FLR,DF},  
but can be solved  analytically.
\bibitem{PIN} The equations of motion Eq(1), Eq(2)  
have  inifinitely  many static solution. We focus on pinned 
states reached  first from flat initial configurations in the 
presence of  time independent external field.
\end{thebibliography}
\end{document}